\documentclass[conference]{IEEEtran}
\IEEEoverridecommandlockouts
\usepackage{cite}
\usepackage{amsmath,amssymb,amsfonts}
\usepackage{algorithmic}
\usepackage{graphicx}
\usepackage{textcomp}
\usepackage{xcolor}
\usepackage{subcaption}
\usepackage{booktabs}
\def\BibTeX{{\rm B\kern-.05em{\sc i\kern-.025em b}\kern-.08em
    T\kern-.1667em\lower.7ex\hbox{E}\kern-.125emX}}
\begin{document}

\title{Outcome Over Learning? Evidence from Student Use of AI Code Assistants for Unit Test Writing
}

\author{\IEEEauthorblockN{Daniel Hinbest}
\IEEEauthorblockA{\textit{Ontario Tech University} \\
\textit{Oshawa, ON, Canada}\\
daniel.hinbest@ontariotechu.net}
\and
\IEEEauthorblockN{Jeremy S. Bradbury}
\IEEEauthorblockA{\textit{Ontario Tech University} \\
\textit{Oshawa, ON, Canada}\\
jeremy.bradbury@ontariotechu.ca}
\and
\IEEEauthorblockN{Michael A. Miljanovic}
\IEEEauthorblockA{\textit{Ontario Tech University} \\
\textit{Oshawa, ON, Canada}\\
michael.miljanovic@ontariotechu.ca}
}

\maketitle

\begin{abstract}
Artificial intelligence (AI) code assistants are rapidly being adopted in programming education, yet little is known about how students naturally use these tools when given unrestricted access. In a user study investigating AI-assisted unit test writing, we observed that a majority of students adopted a one-shot interaction strategy, asking the AI to generate a complete unit test directly from the initial requirements. While this approach is capable of efficiently producing a solution, it shifts students' focus from learning how to write effective unit tests to simply completing the task. Moreover, because students rarely engaged in critically evaluating, refining, or iterating on AI-generated output, they may be missing opportunities to develop the critical thinking skills that are essential for effective and responsible use of AI. In an effort to shift student focus from outcome to learning we also discuss how training, accountability and improved learning task design can better support a more learning-centric responsible use of AI.
\end{abstract}

\begin{IEEEkeywords}
Generative AI, Large Language Models, undergraduate studies, unit tests, computer science education
\end{IEEEkeywords}

\section{Introduction}
Generative Artificial Intelligence (GenAI) is changing Computer Science (CS) education at an unprecedented rate.
Major academic institutions, including Harvard University, are rapidly adopting various forms of GenAI in their CS programs~\cite{liu2024teaching}, while the current widespread availability and free access of GenAI chatbots and tools has led to high usage by students. 

GenAI has quickly become a disruptive force in higher education, highlighting the problems associated with the traditional CS evaluation of product over process. Institutions have been woefully under-prepared for the arrival of this technology, which has highlighted the flaws and limitations of the ways students are assessed~\cite{alier2024generative}. A Large Language Model (LLM) can write an essay or solve a coding exercise in seconds -- tasks that would typically take students hours. Although students have jumped at the opportunity to complete their coursework exercises more quickly than ever, over-reliance on GenAI has already been shown to have a significant negative correlation between its usage and critical thinking~\cite{gerlich2025ai, BBS+25, LST+25}. 

Despite the potential harms of this technology, including both educational and environmental concerns, many researchers argue that the usage of GenAI in educational institutions should not be prohibited~\cite{choi2023chatting, garcia2023new, iskender2023holy, lim2023generative, perkins2023academic, tlili2023if}. Students have easy access to LLMs outside of their educational institutions, and are unlikely to rely exclusively on proprietary AI tools as long as free or low-cost alternatives exist. Furthermore, students will be expected to interact with such systems after completion of their studies~\cite{masters2023ethical}, and those who refuse to engage at all with the technology potentially face a significant disadvantage.

Unfortunately, students who are relying too much on GenAI to complete their assignments are setting themselves up to be fully unprepared for the world after graduation. In an ideal scenario, students could leverage the capabilities of GenAI to improve their learning, treating it as a 24/7 fully personalized mentor that can answer most CS1 questions and solve problems with remarkable accuracy~\cite{dwivedi2023opinion}. In practice, this behaviour is rare and only observed by students who are already actively engaged and motivated to learn. Disconcertingly, many students instead use GenAI as a problem-solver, offloading their critical thinking rather than spending the time and effort needed to develop their skills. 

We take the position that students' use of AI code assistants is beneficial and should be guided, through training, accountability, and task design. Much of the prior work work assesses how students perceive and use GenAI in computer science education, but little of it observes student use directly. One particular relevant prior study focused on novice Python programmer programming without AI and with AI via Github Copilot. In their study that noted that programming with AI increased performance (both actual and perceived) and no differences in emotion and self-efficacy were observed~\cite{GPR24}. While learning wasn't directly studied the authors did recommend educators to \emph{``...consider the importance of experiential learning with [Artificial Intelligence-driven Development Environments] AIDEs and their potential to help struggling students build self-efficacy wit programming via new skillsets"}~\cite{GPR24}.

Our work 
builds on this prior work and explores how students write unit tests with and without the use of an AIDE (e.g., IDE with an AI code-assistant plugin). This contribution informs a series of best practices to support CS educators adapt their course AI policy and assessments to encourage critical use of AI as a means to support learning. 
Section 2 explores related work on applications of GenAI in computer science education and student use of GenAI. Section 3 outlines our methodology, including the study procedures and analysis of the AI use. Our findings from the activities are discussed in Section 4, with a discussion around the findings in Section 5. Section 6 includes our recommendations for best practices for the use of AI in software testing.

\section{Related Work}

We organize the related work into two areas: efforts to understand the use of GenAI in CS education, and studies specifically focused on student interactions with AI-based code assistants in their course activities.

\subsection{Use of GenAI in Computer Science Education}

Integrating an AI-based smart assistant into the testing game Code Defenders, Mezzaro et al. found students expected AI to provide useful explanations and code during the interactions. While the smart assistant was able to provide useful responses, the test cases were not useful and in some cases broken. As a result, the students did not benefit from the smart assistant \cite{mezzaro2024empirical}. This study examined an assistant built into a testing game, whereas we examine how students use an existing coding assistant on assigned unit testing tasks.

Research in software testing education on guided LLM use with instructor guidance or iterative prompt refinement has been shown to improve students' critical evaluation of AI-generated artifacts and produce stronger learning outcomes. Additionally, unguided use has often led to superficial adoption or blind trust \cite{dwarakanath2026insights}. Our work adds to this by exploring further into what students do when given unrestricted use to AI tools.

Fernandez and Nelson had students record and compare their programming sessions without and with GenAI, then reflect on the recordings using a structured framework. The analysis identified several themes occurring: students learned to slow down and understand code before they write or generate it, recognized patterns in their own approaches for problem solving, and process improvements. They also reflected on AI limitations and developed strategies for critical use, including improved prompting and use to support learning rather than complete tasks \cite{fernandez2026reflection}. Our work builds on this by exploring the chat logs themselves, adding further insight on student experiences in AI with unit testing.

Taken together, these studies examine purpose-built assistants, guidance strategies, and students’ reflections. They do not show what students ask an AI to do, or how they interact with it, when they use an existing coding assistant on unit testing tasks.

\subsection{Use of GenAI by Students}

Research has shown students most often use GenAI to summarize material and support personalized learning. While GenAI users performed better on the specific question posed, non-users performed better overall, suggesting reliance on GenAI may act as a shortcut that weakens problem-solving strategies and does not support deeper learning over time; the authors recommend educators guide students toward more effective AI use \cite{osorio2025impact}. These are useful general course findings but do not focus in testing activities.

Birillo et al. studied first and second year students using an AI-based hint system offering tiered textual and code hints integrated into an in-IDE plugin to learn Kotlin. They found that most hint interactions were positive or neutral, with most negative cases resulting from system errors rather than hint quality, and that students often resolved unhelpful hints with multiple requests and combining suggestions \cite{birillo2026strategies}. Our work adds to this by exploring how students use open-ended assistants in the unit testing process.

Wood and Moss evaluated GenAI's impact on learning experiences and perceptions in a master's level instructional design course using pre- and post-course surveys, reflective assignments, and questionnaires. Their work found that students' use of GenAI increased their comfort with the technology and their understanding of its ethical impacts. They offer several recommendations such as gradual integration with clear learning objectives, preserving the role of human instruction, providing opportunities for use in real-world scenarios, and encouraging student feedback and discussion around the use of GenAI \cite{wood2024evaluating}. This work relied on surveys and reflections in a non-testing course, but does not explore interactions themselves or undergraduate experiences in writing tests.

Prather et al. studied novice programmers use of GenAI tools and observed participants in lab session with eye tracking as well as conducted interviews~\cite{PRL+24}. Their study results found students already positions to be successful benefited from GenAI while students who struggled could experience a negative impact including an illusion of competence.

Most of this work concerns general course use, a purpose-built tool, or students’ own reports of their use, and little of it concerns unit testing. We address this gap by analyzing students’ chat logs from an AI coding assistant on two unit testing tasks, to characterize what they ask for, what they provide, and how often they check the result.

\section{Methodology}


To understand how students utilized AI code assistants to write unit tests, we designed a study consisting of a survey and a set of unit testing tasks -- some completed without the use of AI and some with AI. The study participants were all CS students enrolled in a third-year undergraduate software quality assurance course. Unit testing was part of the course curriculum, and students had already learned about unit testing and completed a hands-on lab activity prior to the study. 
Study participation was voluntary, and participants received a bonus mark toward their course grade. In total, 32 students from the course agreed to participate in the study\footnote{The study was approved by the Ontario Tech Research Ethics Board ($\#$18160) and all participants reviewed and signed a consent form prior to beginning the study.} and after excluding students who did not complete the study tasks or submitted incomplete or incorrect data we had 21 student submissions to analyze. 


First, all participants completed a survey that included questions about basic demographics (gender, year of study), computer science experience (e,g., industry experience, coding experience, testing experience, confidence writing unit tests), and AI tool experience (e.g., AI tools for ).


Next, participants completed four unit testing tasks in Python. For two of the tasks, they were permitted to use an AI code assistant and for two tasks they were not. In both treatments, one task required statement coverage and the other required branch and loop coverage. 

During the AI unit testing tasks, participants were instructed to complete the activities with the use of the AI coding assistant Continue~\cite{continue}, available as a plugin in the  PyCharm IDE. Continue was setup to use gpt4o or gpt5 via an OpenAI API key. At the completion of each task, students were instructed to upload their completed test source files and Continue chat log files. 

Participants were randomly assigned to one of two groups that completed the tasks in opposite orders. Group 1 completed the non-AI tasks first and the AI-assisted tasks second, and Group 2 completed them in the reverse order.

The tasks students were required to complete with the use of AI were:
\begin{itemize}
    \item \textit{Classify Triangle (Statement Coverage)}. A small function that takes three numbers as parameters, and determines whether it is valid and returns invalid if it is not. If it is a valid triangle, it identifies a triangle as equilateral, isosceles, or scalene based on how many sides are equal.
    \item \textit{Process Cart (Branch \& Loop Coverage)}. A function that takes in two parameters. The first one is the items, as a list of dictionaries. The second, an optional parameter is a coupon. This function calculates the prices and quantity of items in a cart with the use of a for loop, and applies coupons to return a final cost. 
\end{itemize}


\subsection{Data Exclusions}
\label{sec:exclusions}
In total, 32 students participated in the study. Although we made significant efforts to analyze all submissions, some data had to be excluded for one of two reasons. The first was incomplete data, such as missing test files or Continue logs, which would compromise the quality of the results. The second was non-relevant submissions, where the uploaded files were not related to the assigned unit testing task. In total ten participants were excluded for incomplete data and one for irrelevant submissions, leaving 21 participants with valid data. After data exclusions the remaining 21 participants included 12 from Group 1 and 9 from Group 2. 

\subsection{Data Analysis}

 This paper focuses only on an analysis of the the AI-assisted tasks and combines the survey data and the Continue chat logs. In the case of the survey data we analyzed the basic demographics, experiences in programming and testing, and their AI experience. In the case of the Continue chat logs, we analyzed both the participant messages and data sent to the chat bot as well as the chat bot responses. In the future we plan to report the full study results include the non-AI tasks results.

\subsubsection{Survey Responses}
While most of the questions were in a closed format, the open-ended question "How do you usually use AI coding assistants?" required a manual analysis process where the responses were sorted by all three authors into nine categories based on the types of responses observed: Debugging \& Problem Solving, Code Review, Improvement \& Optimization, Starter Code Generation, Code Block Generation \& Auto-completion, Learning, Testing, General Development, and Limited/No AI use. An individual response could be sorted into multiple categories based on the author coding and discussion.

\subsubsection{AI Interactions}
Using the Continue AI chat log, we reviewed both the participant's chat messages and the AI responses to understand their interaction during the unit testing activities. 
To understand the patterns observed in each unit testing task we assessed the logs to determine: if assistant and/or agent mode was used, what types of prompts were used, and what supplemental information/files was provided to support the prompt requests. With the exception of the prompt type and prompt content coding which were completed by all three authors, all other data was analyzed by author one. Specifically, each chat log for each tasks was analyzes to identify:

\begin{itemize}
    \item \textbf{AI Mode.} The Continue AI code assistant in PyCharm can be run in one of two modes: agent or assistant. Agent mode uses AI to generate code that is directly added to the output file without human intervention. Assistant mode requires programmer approval to add generated code to an output file. In Assistant mode, the programmer often asks the AI chatbot questions or gets feedback on their work. It is possible to have instances where a user may apply both in the same task. Both modes were available to participants and we did not direct them to use a specific AI mode. 
    \item \textbf{AI Interaction.} To understand the interaction between the participant and the AI, we first look at the number of prompts in the chat log -- whether it was a one-shot approach or more conversational. One-shot interactions are defined typically as a single prompt or potentially as a couple of prompts that were independent and asking different tasks to be complete. Conversational interactions are when participants interacted with the AI as if they were having a conversation in which each prompt/reply interaction naturally flowed from the previous interaction. In many cases, conversational interactions are longer and are composed of sequences of prompt/reply pairs. 
    \item \textbf{Prompt Types.} To understand the purpose of a given prompt or sequence of prompts we categorize the interaction by prompt type:
    \begin{itemize}
        \item \emph{Process help:} guidance on the process of writing unit tests or reaching code coverage
        \item \emph{Problem clarifications:} prompts that request explanation of the problem or task, what the source code under test does, what the tests are supposed to do, or explanation of problem/task aspects such as test coverage definitions.
        \item \emph{Code generation:} prompts that either request that completely or partial generate the requested test case.
        \item \emph{Debugging:} prompts focused on fixing identified bugs in the code under test or the tests themselves.
        \item \emph{Validation:} prompts which aim to evaluate the AI output from previous prompts and validate that it is correct. For example, a participant may ask whether a test is correct or if a test suite meets the expected coverage requirements.
    \end{itemize}
    \item \textbf{Prompt Content.} To understand how prompts are composed, we consider what information and supporting artifacts is provided to the AI code assistant. Specifically we consider the following information and supporting artifacts: 
    \begin{itemize}
        \item \emph{Task instructions:} The prompt can include task instructions in one of the following forms: verbatim, a paraphrasing of the instructions given, or an enhanced representation of the task instruction given which included additional details.
        \item \emph{Coverage definitions:} The prompt can include a definition of the code coverage required for a given unit test. 
        \item \emph{Uploaded source file:} The prompt can include source code files of the software under test as context. 
        \item \emph{Uploaded test files:} The prompt can include test files as context. A test file can be either partially completed or an empty one.
        \item \emph{Source code:} The prompt can also include source code embedded directly in the prompt text. This is similar to uploading the source code as a file
        \item \emph{Question:} The prompt can include questions to the AI, such as verifying the participant's understanding of the unit testing task as well as asking for help breaking down the task into sub-tasks.
        \item \emph{Program output:} The prompt can also include program and test outputs as well as error messages in the prompt text.
    \end{itemize}
\end{itemize}



\section{Results}



\subsection{Survey Responses}

\begin{table}[t!]
\centering
\caption{On a scale of 1-5, how confident do you feel about writing unit tests?}
\label{tab:confidence}
\begin{tabular}{lrr}
\toprule
\textbf{Answer} & \textbf{Count} & \textbf{Percentage} \\
\midrule
1 - Not confident at all & 3 & 14.3\% \\
2 & 9 & 42.8\% \\
3 - Moderately confident & 8 & 38.1\% \\
4 & 1 & 4.8\% \\
5 - Extremely confident & 0 & 0.0\% \\
\bottomrule
\end{tabular}
\end{table}
 
\begin{table}[t!]
\centering
\caption{Have you used AI assistance in testing before?}
\label{tab:ai_experience}
\begin{tabular}{lrr}
\toprule
\textbf{Answer} & \textbf{Count} & \textbf{Percentage} \\
\midrule
Yes & 17 & 81.0\% \\
No & 4 & 19.0\% \\
\bottomrule
\end{tabular}
\end{table}

The survey responses presented are based on the 21 participants with valid data after exclusions (see Section~\ref{sec:exclusions} for details).  All students were upper-year CS students enrolled in a third year quality assurance class - 16 identified as third-year students and 5 identified as fourth-year students. 
Four of the students reported industry experience through an internship, co-operative education or prior work experience with only one student indicating experience writing unit tests as part of their work experience. All students had previously learned about unit testing in the course and completed a hands-on lab activity. To understand their perceived ability to write unit tests we asked them to rate their confidence. Table~\ref{tab:confidence} presents the confidence in unit testing writing data with many of the students reporting low-to-moderate confidence or being moderately confident. Fewer students expressed limited or higher levels of confidence in writing unit tests.

\begin{table}[t!]
\centering
\caption{Which of these AI models have you used before?}
\label{tab:model}
\begin{tabular}{lrr}
\toprule
\textbf{Model} & \textbf{Count} & \textbf{Percentage} \\
\midrule
GPT & 20 & 95.2\% \\
Gemini & 14 & 66.7\% \\
Claude & 10 & 47.6\% \\
Copilot & 9 & 42.9\% \\
Llama & 2 & 9.5\% \\
Cursor & 1 & 4.8\% \\
V0 & 1 & 4.8\% \\
\bottomrule
\end{tabular}
\end{table}
 
\begin{table}[t!]
\centering
\caption{How confident are you in using AI coding assistants?}
\label{tab:ai_confidence}
\begin{tabular}{lrr}
\toprule
\textbf{Answer} & \textbf{Count} & \textbf{Percentage} \\
\midrule
1 - Not confident at all & 0 & 0.0\% \\
2 & 2 & 9.5\% \\
3 - Moderately confident & 8 & 38.1\% \\
4 & 7 & 33.3\% \\
5 - Extremely confident & 4 & 19.1\% \\
\bottomrule
\end{tabular}
\end{table}

Students were also asked about their experiences using AI code assistants (or chatbot) in their programming and testing activities. Of the 21 students whose submissions were used in the analysis, 17 (80.9\%) reported having used AI code assistant in their testing activities (see Table~\ref{tab:ai_experience}). When asked about which AI models they had used to aid them in coding or testing the most popular choice was the use of OpenAI's GPT models which had been used by 20 (95.2\%) of the students (see Table~\ref{tab:model}). Google's Gemini was the next most used followed by Anthropic's Claude and Microsoft's Copilot.


Students also reported moderate-to-high confidence in using AI assistants to complete their tasks (see Table~\ref{tab:ai_confidence}) with two students reporting low-to-moderate confidence, eight reporting moderately confident, seven reporting moderate-to-high confidence, and four reporting high confidence. An interesting note is that the students expressed higher confidence in AI code assistants, for which that had received no formal training in their courses, thank they did with unit testing for which they had formally learned and practiced in the weeks leading up to the study. In other words, students were far less confident in writing unit tests (20 of 21 rated themselves at 3 or lower) than in using AI code assistants (11 of 21 rated themselves 4 or 5). These students were therefore familiar with AI tools but not confident in the skill the task was meant to assess, which may make delegating the work to AI appealing.

\begin{table}[t!]
\centering
\caption{How do you usually use AI coding assistants? (Category Totals)}
\label{tab:ai_usage}
\begin{tabular}{lrr}
\toprule
\textbf{Use Case} & \textbf{\# Students} & \textbf{\% of Students} \\
\midrule
Debugging \& Problem Solving & 10 & 47.6\% \\
Learning & 6 & 28.6\% \\
Starter Code Generation & 5 & 23.8\% \\
Improvement \& Optimization & 4 & 19.1\% \\
Code Review & 3 & 14.3\% \\
Code Block Generation & 3 & 14.3\% \\
Full Code Generation & 3 & 14.3\% \\
General Development & 2 & 9.5\% \\
Limited to no AI use & 1 & 4.8\% \\
Testing & 1 & 4.8\% \\
\bottomrule
\end{tabular}
\end{table}

In addition to understanding students experience with different AI code assistants and their confidence in using AI code assistants we were also interested in their usage patterns. This question was open-ended and we reviewed all of the responses and categorizes the usage into ten different categories. Table~\ref{tab:ai_usage} presents the self-reported usage of AI for coding and testing tasks. The most popular tasks AI was used to support were debugging and problem solving (47.6\%), learning (28.6\%) and starter code generation (23.8\%). Interestingly for our task only one student (4.8\%) indicated they usually used an AI code assistant for testing even though 17 of 21 students indicated that they had used an AI code assistant for testing. This indicates that although the majority had tried testing with an AI code assistant it was not a normal development practice. It was also interesting to observe that only 14.3\% of students reported using AI for full code generation which we see later is inconsistent with what we observed in practice. Lastly, one student wrote that they limit their use of AI, responding: \emph{"I do not use them actively, if they are integrated with my IDE then at most they would simplify the process of typing out code by showing a suggested prediction of what I meant to type and could autofill."}

\subsection{Task 1: Statement Coverage for Classify Triangle}

The Continue coding results for the Classify Triangle task are shown in Figures 1 through 4. As shown in Figure 1, 57\% of students only used the assistant, 29\% used only the agent, and 14\% used both at some point. Figure 2 shows that 76\% of interactions were one-shot, meaning students gave directions without building on previous prompts, while 24\% were conversational. Students averaged two prompts, with one outlier writing 8.

Figure 3 shows the prompt types. Nearly all students (90\%) asked the AI to complete code, and 24\% asked it to validate tests. Clarifications about test cases and the purpose of the activity (e.g., coverage definitions) were less common, at 14\% and 10\% respectively.

Figure 4 shows what students provided in their prompts. Test definitions were the most common (57\%), followed by source code, which 52\% uploaded and 33\% pasted into the chat. Many students also included some form of instructions, either verbatim, which 38\% did or paraphrased as was done in 33\% of submissions. 10\% of students added detail to the instructions and another 10\% uploaded an empty test file.

\begin{figure}[htbp]
\centering
\includegraphics[width=1\linewidth]{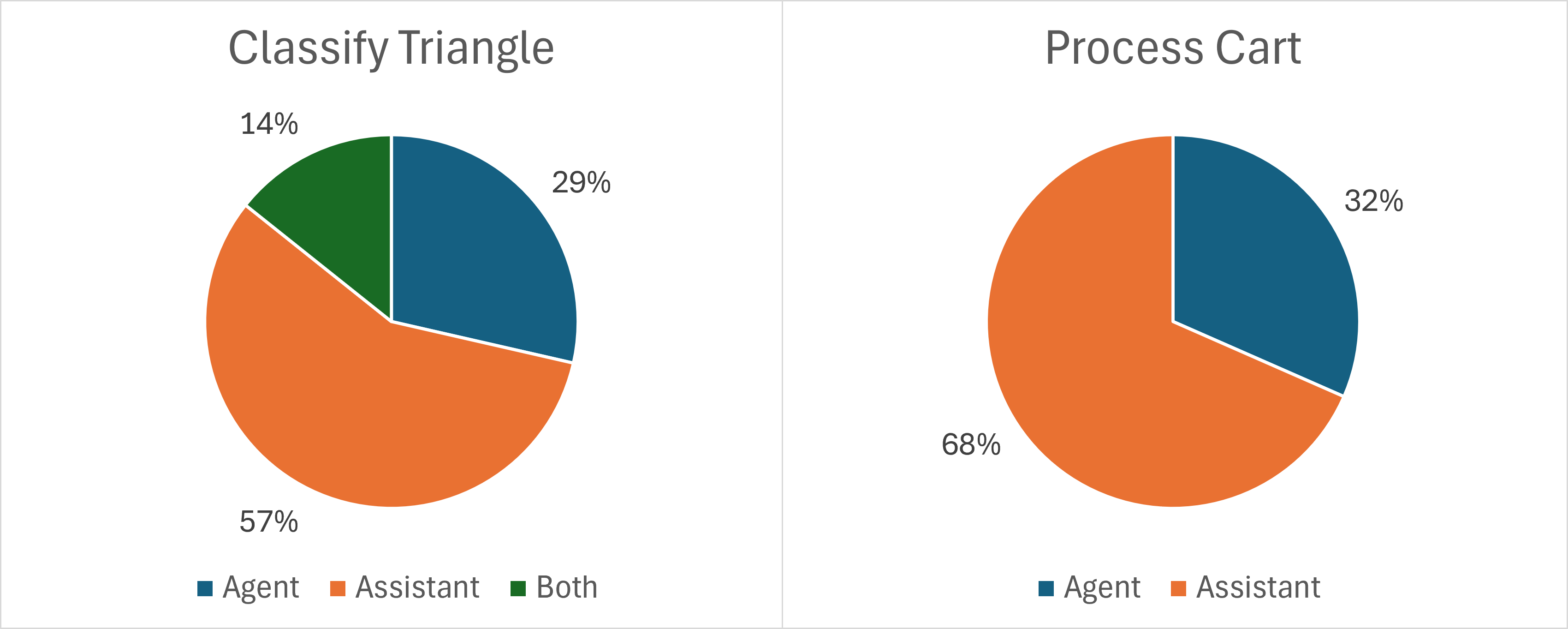}
\caption{AI Mode}
\label{fig:mode} 
\end{figure}

\begin{figure}[htbp]
\centering
\includegraphics[width=1\linewidth]{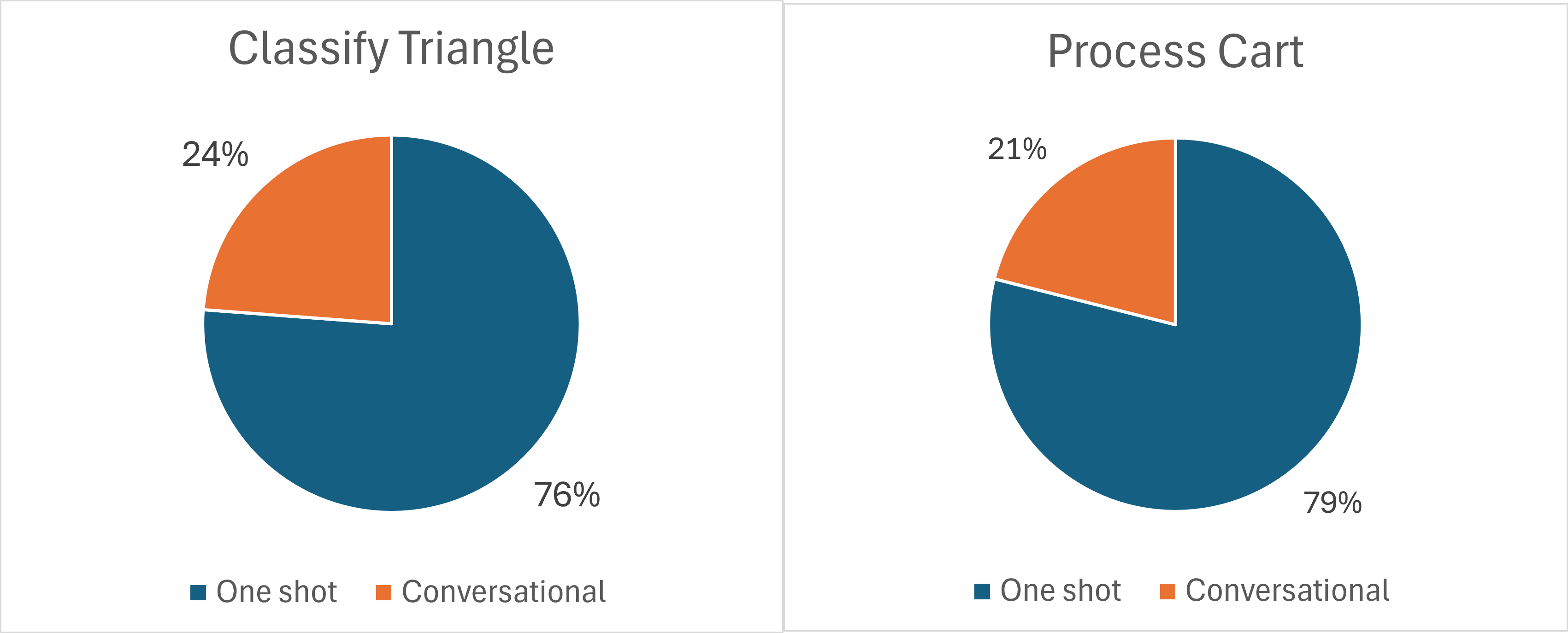}
\caption{AI Interaction}
\label{fig:interactions} 
\end{figure}

\begin{figure}[htbp]
\centering
\includegraphics[width=\linewidth]{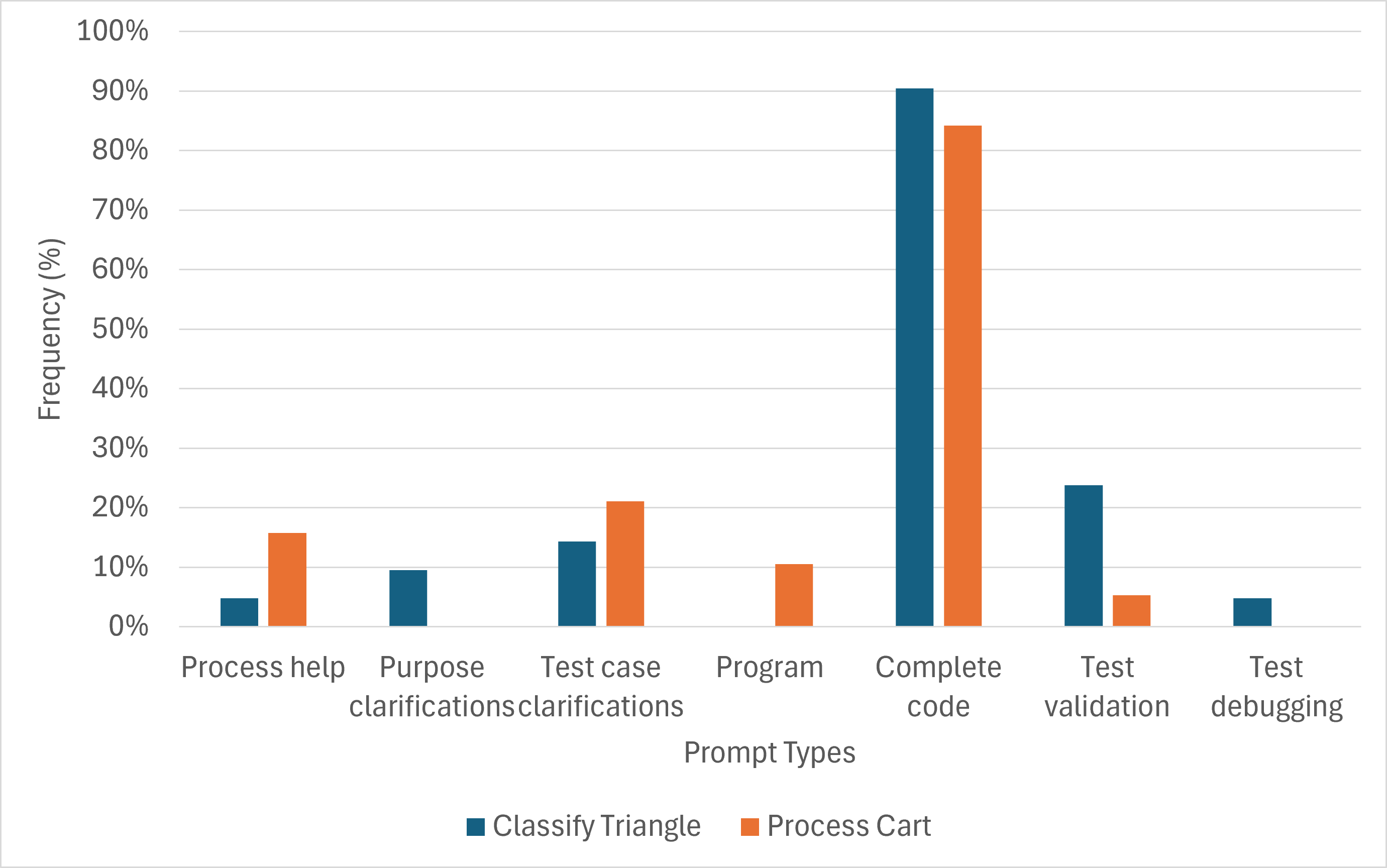}
\caption{Prompt Type}
\label{fig:prompt_type} 
\end{figure}

\begin{figure*}[t!]
\centering
\includegraphics[width=0.95\textwidth, height=6.5cm, keepaspectratio=false]{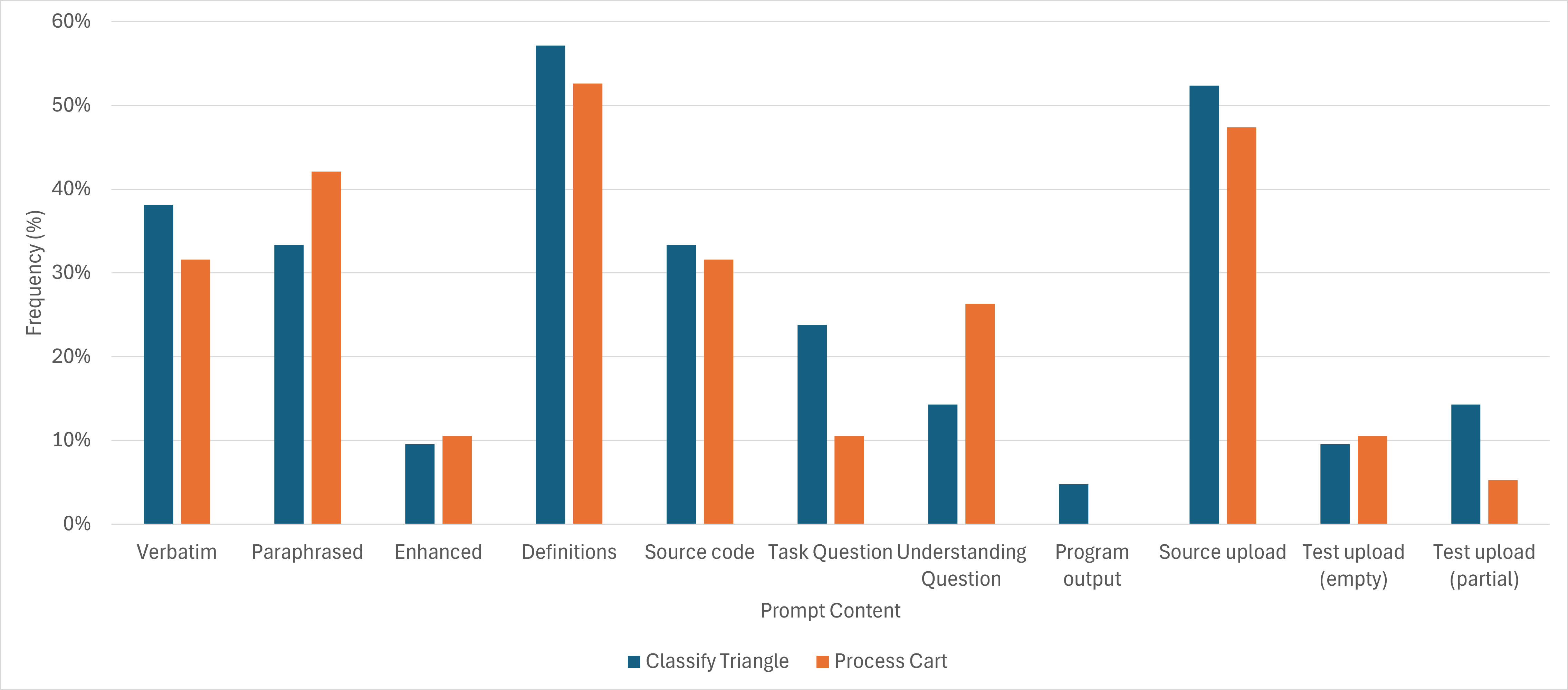}
\caption{Prompt Content}
\label{fig:prompt_content} 
\end{figure*}

\subsection{Task 2: Branch \& Loop Coverage for Process Cart}


Branch and loop coverage was the second AI-related activity, and 19 of 21 students completed it. Unlike the first task, no students changed between agent and assistant mode with 68\% using the assistant and the remaining 32\% using the agent (Figure 1). Similar to the first task, most interactions were again one-shot (79\%), and students averaged 2.7 prompts, with one outlier exchanging 21 prompts (Figure 2).

Complete code remained the most common prompt type (84\%) as shown in Figure 3. Compared with Classify Triangle, fewer students asked the AI to validate their tests (5\% vs 24\%), and there was an increase in clarification about test cases at 21\%, the program at 11\%, or the testing process at 16\%.

Figure 4 shows that students again shared the task code (47\% uploading it and 32\% pasting it) and instructions (42\% paraphrased, 32\% verbatim, and 11\% enhanced), with 53\% providing definitions. Understanding questions were more common than in the first task, increasing to 26\% from 14\%, and questions about the task were less common - dropping to 11\%, down from 24\%. 11\% uploaded an empty test file and 5\% uploaded a partially complete one.

\section{Discussion}
Our results show that students often used AI coding assistants not to assist them in completing the task but to do the task for them. Most students asked the AI to generate the full test code (90\% in Classify Triangle and 84\% in Process Cart), and few asked it to validate tests, whether written by themselves or the AI (24\% and 5\%). This could indicate trust in the AI's output, although it wasn't measured directly. Students also showed comfort and confidence when they use AI coding assistants, with many students using them for debugging, problem-solving, or learning. However, they were less confident in their own development of unit tests, even though most of them had used AI to assist in testing previously.

Across both tasks, students used Continue in a consistent way, with roughly three quarters of interactions being one-shot. The shift to branch and loop coverage is notable. Students asked more questions about the program and the testing process, yet fewer asked the AI to validate their tests (24\% to 5\%). If students mostly request complete solutions and rarely ask for feedback on their own tests, training, accountability, and task design need to steer them toward the second use, which is the basis for the recommendations in Section 7.

These findings suggest students who lack confidence in unit test development may use AI to support them or to complete tests for them. This paper did not analyze learning outcomes, test quality, or time spent on the task, so we cannot say if this helps or hinders learning. It is possible that relying on AI to complete tasks limits the practice where students learn to write tests and build confidence, which is a question for future work. Educators should consider guiding students to use these tools to support their learning.

\section{Limitations}

This study has several limitations. First, our sample is small and comes from a single upper-year undergraduate software quality assurance course at one institution, with 21 of 139 enrolled students included in the analysis. Participation was voluntary and students received a grade bonus, so the participants may differ from the rest of the class, and our findings may not generalize to other courses, levels, or institutions. Second, 11 of the 32 students who volunteered (34\%) were excluded, mostly because of incomplete data. We could identify the group for only 5 of them, so we cannot rule out that the exclusions affected the remaining sample. Third, this paper does not analyze learning outcomes, test quality, or time on task, and we did not measure trust directly, so our conclusions about learning and critical thinking are interpretations of student behaviour and not measured effects.

\section{Best Practices to Support Learning}


We believe the behaviours we observed in the use of AI code assistants for writing unit tests are concerning from a learning perspective. The challenge is how do we mitigate negative behaviours and promote positive behaviours that encourage learning and not simply outcome achievement. We advocate that developing better learner behaviours in LLM usage requires a combination of training, accountability, and improved task design that includes context. 

First, with respect to training we see two viable approaches:

\begin{itemize}
\item \emph{Upfront training:} provide learners with upfront training on how to effectively use AI code assistants and LLMs to promote critical thinking as well as educate learners on the importance of critical thinking as a necessary skills in computer science. 
\item \emph{Continuous training via prompt feedback:} designing custom AI Code Assistants that by default provide advice on how to improve prompts with respect to learning in addition to answering learners questions and completing learners requested tasks.
\end{itemize}

Our data supports both approaches: most students (17 of 21) had used AI for testing before, yet few prompts involved checking the output, and only 21 to 24\% of interactions were conversational, so a feedback tool often has a single prompt to act on.

Upfront training on LLM usage and designing custom AI code assistants that provide feedback on best practices for LLM usage are both ways of encouraging better practices that promote learning. However, both rely on learners' self motivation and interest in learning to be successful. 

To support good LLM practices we also advocate building accountability around LLM usage into learning tasks. One way to achieve this is to require the submission of AI code assistant chat history along with the learning task deliverables like test cases. This requirement can be combined with marking rubrics that assign value to using LLMs to support learning and not just produce deliverables. To be truly effective, the best practices identified in the rubric need to be clearly defined via training. 

Lastly, improved learning task design that accounts for LLM usage and pitfalls is necessary. For example, requesting more than the final deliverable and instead requesting students provide supplemental information to demonstrate understanding. Learning task design can be further strengthened by embedding individual tasks into larger projects that require more than just code generation. In our study, many students gave the AI the task instructions and the source code, and most asked for complete code, so a task that can be pasted is likely to be completed. In Classify Triangle, for example, test creation could be only one part of the task: before writing tests, students would first determine which inputs are needed to cover each statement.

Supporting accountability via submission of chat logs and improving the task design of learning activities with LLMs are both methods that increase the difficulty associated with using LLMs to achieve outcomes without learning. They can still be circumvented but reduce the workload difference between using LLMs to simply complete tasks and using LLMs to learn. 

One direction we don't advocate using is limiting LLM capabilities and restricting LLM chat functionality. With free alternatives, students may either not use the limited LLM offering or will use it with support from an unrestricted LLM. 

\section{Conclusions}

Based on our observations of student behaviour, it is clear that there is a critical need to take action regarding student use of generative AI in the context of software testing. This does not mean that we recommend the complete prohibition of LLMs in undergraduate CS education, which is likely to be an ineffective strategy. Not only will students continue to use these tools (secretly, rather than with proper instructor guidance), but students will be expected to be familiar and experienced with these tools at the completion of their degree. Although there is great potential in the use of LLMs in CS education, this can only be achieved through responsible practices. The unfettered use of LLMs by students has significant risks, not only developmentally but also ethically and environmentally. It is our responsibility to guide our students and ensure they are ethical users of AI in today's world.

\bibliographystyle{IEEEtran}
\bibliography{references}

@inproceedings{LST+25,
author = {Lee, Hao-Ping (Hank) and Sarkar, Advait and Tankelevitch, Lev and Drosos, Ian and Rintel, Sean and Banks, Richard and Wilson, Nicholas},
title = {The Impact of Generative AI on Critical Thinking: Self-Reported Reductions in Cognitive Effort and Confidence Effects From a Survey of Knowledge Workers},
year = {2025},
isbn = {9798400713941},
publisher = {Association for Computing Machinery},
address = {New York, NY, USA},
url = {https://doi.org/10.1145/3706598.3713778},
doi = {10.1145/3706598.3713778},
booktitle = {Proceedings of the 2025 CHI Conference on Human Factors in Computing Systems},
articleno = {1121},
numpages = {22},
location = {Yokohama, Japan},
series = {CHI '25}
}

@article{BBS+25,
author = {Hamsa Bastani  and Osbert Bastani  and Alp Sungu  and Haosen Ge  and Özge Kabakcı  and Rei Mariman },
title = {Generative AI without guardrails can harm learning: Evidence from high school mathematics},
journal = {Proceedings of the National Academy of Sciences},
volume = {122},
number = {26},
pages = {e2422633122},
year = {2025},
doi = {10.1073/pnas.2422633122},
URL = {https://www.pnas.org/doi/abs/10.1073/pnas.2422633122},
eprint = {https://www.pnas.org/doi/pdf/10.1073/pnas.2422633122}}

@inproceedings{PRL+24,
author = {Prather, James and Reeves, Brent N and Leinonen, Juho and MacNeil, Stephen and Randrianasolo, Arisoa S and Becker, Brett A. and Kimmel, Bailey and Wright, Jared and Briggs, Ben},
title = {The Widening Gap: The Benefits and Harms of Generative AI for Novice Programmers},
year = {2024},
isbn = {9798400704758},
publisher = {Association for Computing Machinery},
address = {New York, NY, USA},
url = {https://doi.org/10.1145/3632620.3671116},
doi = {10.1145/3632620.3671116},
booktitle = {Proceedings of the 2024 ACM Conference on International Computing Education Research - Volume 1},
pages = {469–486},
numpages = {18},
location = {Melbourne, VIC, Australia},
series = {ICER '24}
}

@inproceedings{GPR24,
author = {Gardella, Nicholas and Pettit, Raymond and Riggs, Sara L.},
title = {Performance, Workload, Emotion, and Self-Efficacy of Novice Programmers Using AI Code Generation},
year = {2024},
isbn = {9798400706004},
publisher = {Association for Computing Machinery},
address = {New York, NY, USA},
url = {https://doi-org.uproxy.library.dc-uoit.ca/10.1145/3649217.3653615},
doi = {10.1145/3649217.3653615},
booktitle = {Proceedings of the 2024 on Innovation and Technology in Computer Science Education V. 1},
pages = {290–296},
numpages = {7},
location = {Milan, Italy},
series = {ITiCSE 2024}
}

@online{continue,
  title={Continue},
  url={https://www.continue.dev/},
  note={Accessed: June 18, 2026}
}

@article{alier2024generative,
  title={Generative artificial intelligence in education: From deceptive to disruptive},
  author={Alier Forment, Marc and Garc{\'\i}a Pe{\~n}alvo, Francisco Jos{\'e} and Camba, Jorge D},
  journal={International Journal of interactive multimedia and artificial intelligence},
  volume={8},
  number={5},
  pages={5--14},
  year={2024}
}

@article{dwivedi2023opinion,
  title={Opinion Paper:“So what if ChatGPT wrote it?” Multidisciplinary perspectives on opportunities, challenges and implications of generative conversational AI for research, practice and policy},
  author={Dwivedi, Yogesh K and Kshetri, Nir and Hughes, Laurie and Slade, Emma Louise and Jeyaraj, Anand and Kar, Arpan Kumar and Baabdullah, Abdullah M and Koohang, Alex and Raghavan, Vishnupriya and Ahuja, Manju and others},
  journal={International journal of information management},
  volume={71},
  pages={102642},
  year={2023},
  publisher={Elsevier}
}

@article{masters2023ethical,
  title={Ethical use of artificial intelligence in health professions education: AMEE Guide No. 158},
  author={Masters, Ken},
  journal={Medical teacher},
  volume={45},
  number={6},
  pages={574--584},
  year={2023},
  publisher={Taylor \& Francis}
}

@article{tlili2023if,
  title={What if the devil is my guardian angel: ChatGPT as a case study of using chatbots in education},
  author={Tlili, Ahmed and Shehata, Boulus and Adarkwah, Michael Agyemang and Bozkurt, Aras and Hickey, Daniel T and Huang, Ronghuai and Agyemang, Brighter},
  journal={Smart learning environments},
  volume={10},
  number={1},
  pages={15},
  year={2023},
  publisher={Springer}
}

@article{perkins2023academic,
  title={Academic integrity considerations of AI large language models in the post-pandemic era: ChatGPT and beyond},
  author={Perkins, Mike},
  journal={Journal of University Teaching and Learning Practice},
  volume={20},
  number={2},
  pages={1--24},
  year={2023},
  publisher={Open Access Publishing Association (OAPA) Launceston, Tasmania}
}

@article{lim2023generative,
  title={Generative AI and the future of education: Ragnar{\"o}k or reformation? A paradoxical perspective from management educators},
  author={Lim, Weng Marc and Gunasekara, Asanka and Pallant, Jessica Leigh and Pallant, Jason Ian and Pechenkina, Ekaterina},
  journal={The international journal of management education},
  volume={21},
  number={2},
  pages={100790},
  year={2023},
  publisher={Elsevier}
}

@article{iskender2023holy,
  title={Holy or unholy? Interview with open AI’s ChatGPT},
  author={Iskender, Ali},
  journal={European journal of tourism research},
  volume={34},
  pages={3414--3414},
  year={2023}
}

@article{choi2023chatting,
  title={Chatting or cheating? The impacts of ChatGPT and other artificial intelligence language models on nurse education},
  author={Choi, Edmond Pui Hang and Lee, Jung Jae and Ho, Mu-Hsing and Kwok, Jojo Yan Yan and Lok, Kris Yuet Wan},
  journal={Nurse Education Today},
  year={2023},
  publisher={Elsevier}
}

@article{wood2024evaluating,
  title={Evaluating the impact of students' generative AI use in educational contexts},
  author={Wood, Dwayne and Moss, Scott H},
  journal={Journal of Research in Innovative Teaching \& Learning},
  volume={17},
  number={2},
  pages={152--167},
  year={2024},
  publisher={Emerald Publishing Limited}
}

@article{garcia2023new,
  title={The new reality of education in the face of advances in generative artificial intelligence},
  author={Garc{\'\i}a-Pe{\~n}alvo, Francisco J and Llorens Largo, Fara{\'o}n and Vidal, Javier and others},
  year={2023},
  publisher={Asociaci{\'o}n Iberoamericana de Educaci{\'o}n Superior a Distancia (AIESAD)}
}

@inproceedings{mezzaro2024empirical,
  title={An empirical study on how large language models impact software testing learning},
  author={Mezzaro, Simone and Gambi, Alessio and Fraser, Gordon},
  booktitle={Proceedings of the 28th international conference on evaluation and assessment in software engineering},
  pages={555--564},
  year={2024}
}

@inproceedings{liu2024teaching,
  title={Teaching CS50 with AI: leveraging generative artificial intelligence in computer science education},
  author={Liu, Rongxin and Zenke, Carter and Liu, Charlie and Holmes, Andrew and Thornton, Patrick and Malan, David J},
  booktitle={Proceedings of the 55th ACM technical symposium on computer science education V. 1},
  pages={750--756},
  year={2024}
}

@article{gerlich2025ai,
  title={AI tools in society: Impacts on cognitive offloading and the future of critical thinking},
  author={Gerlich, Michael},
  journal={Societies},
  volume={15},
  number={1},
  pages={6},
  year={2025},
  publisher={Multidisciplinary Digital Publishing Institute}
}

@inproceedings{birillo2026strategies,
author = {Birillo, Anastasiia and Rostovskii, Aleksei and Golubev, Yaroslav and Keuning, Hieke},
title = {Understanding Student Interaction with AI-Powered Next-Step Hints: Strategies and Challenges},
year = {2026},
isbn = {9798400722561},
publisher = {Association for Computing Machinery},
address = {New York, NY, USA},
url = {https://doi.org/10.1145/3770762.3772544},
doi = {10.1145/3770762.3772544},
booktitle = {Proceedings of the 57th ACM Technical Symposium on Computer Science Education V.1},
pages = {134–140},
numpages = {7},
location = {USA},
series = {SIGCSE TS 2026}
}

@inproceedings{fernandez2026reflection,
author = {Fernandez, Sarah "Magz" and Nelson, Greg L.},
title = {A 'watch your replay videos' Reflection Assignment on Comparing Programming without versus with Generative AI: Learning about Programming, Critical AI Use and Limitations, and Reflection},
year = {2026},
isbn = {9798400722561},
publisher = {Association for Computing Machinery},
address = {New York, NY, USA},
url = {https://doi.org/10.1145/3770762.3772668},
doi = {10.1145/3770762.3772668},
booktitle = {Proceedings of the 57th ACM Technical Symposium on Computer Science Education V.1},
pages = {323–329},
numpages = {7},
location = {USA},
series = {SIGCSE TS 2026}
}

@inproceedings{dwarakanath2026insights,
author = {Dwarakanath, Samhitha and Nascimento, Nathalia and Guimaraes, Everton},
title = {Software Testing Education in the LLM Era: Insights and Emerging Theory},
year = {2026},
isbn = {9798400724763},
publisher = {Association for Computing Machinery},
address = {New York, NY, USA},
url = {https://doi-org.uproxy.library.dc-uoit.ca/10.1145/3793654.3793757},
doi = {10.1145/3793654.3793757},
booktitle = {Proceedings of the 7th ACM/IEEE International Conference on Automation of Software Test},
pages = {66–70},
numpages = {5},
location = {
},
series = {AST '26}
}

@inproceedings{osorio2025impact,
author = {Ramirez Osorio, Valeria and Zavaleta Bernuy, Angela and Simion, Bogdan and Liut, Michael},
title = {Understanding the Impact of Using Generative AI Tools in a Database Course},
year = {2025},
isbn = {9798400705311},
publisher = {Association for Computing Machinery},
address = {New York, NY, USA},
url = {https://doi.org/10.1145/3641554.3701785},
doi = {10.1145/3641554.3701785},
booktitle = {Proceedings of the 56th ACM Technical Symposium on Computer Science Education V. 1},
pages = {959–965},
numpages = {7},
location = {Pittsburgh, PA, USA},
series = {SIGCSETS 2025}
}

\end{document}